\documentclass[twocolumn,prb,showpacs,preprintnumbers,amsmath,amssymb]{revtex4}
\usepackage{graphicx}
\usepackage{dcolumn}
\usepackage{bm}
\begin{document}

\title{Landau-Zener Interference in Multilevel Superconducting Flux Qubits Driven by Large Amplitude Fields}

\author{ Xueda Wen and Yang Yu}
\email{ yuyang@nju.edu.cn}
\affiliation{National Laboratory of Solid State Microstructures and Department of Physics, Nanjing University, Nanjing 210093, China }

\begin{abstract}
We proposed an analytical model to analyze the Landau-Zener interference in a multilevel superconducting flux qubit driven by large-amplitude external fields.
Our analytical results agree remarkably with those of the experiment [Nature $\mathbf {455}$, 51 (2008)]. Moreover, we studied the effect of driving-frequency
and dephasing rate on the interference. The dephasing generally destroys the interference while increasing frequency rebuilds the interference at large dephasing rate.
At certain driving frequency and dephasing rate, the interference shows some anomalous features as observed in recent experiments.

\end{abstract}

\pacs{74.50.+r, 85.25.Cp}

\maketitle

\begin{displaymath}
 \textbf{I.\ INTRODUCTION }
\end{displaymath}

Superconducting Josephson devices coherently driven by external fields provide new insights into
fundamentals of quantum mechanics and hold promise for use in quantum computation as qubits.\cite{Mooij,Makhlin,Nielsen}
Compared with natural atoms and molecules, these devices can be strongly coupled to external radio-frequency (rf) fields
while preserving quantum coherence.\cite{Chiorescu,Wallraff,Schuster} A large number of experiments associated with
strong driving have been done on these devices such as Rabi oscillations in the multiphoton regime,\cite{Nakamura,Saito,Yu}
Bloch oscillations,\cite{Boulant} Landau-Zener(LZ) interference\cite{Oliver,Sillanpaa,BernsPRL}and dressed states of
superconducting qubits under extreme driving.\cite{Wilson}

Recently, coherent dynamics of superconducting qubits in the regime dominated by LZ transitions were extensively
studied.\cite{Oliver,Sillanpaa,BernsPRL,Shytov,Izmalkov,Izmalkov2,Ankerhold,Ithier,Saito2,Saito3,Shevchenko3}
In this case, the driving frequency is much smaller than energy-level separation and the transitions occur via the LZ process at a level
crossing.\cite{Shytov,Izmalkov} One may use LZ transitions to enhance the quantum tunneling rate,\cite{Ankerhold,Ithier}
to prepare the quantum state,\cite{Saito2} to control the qubit gate operations effectively\cite{Saito3}
and so on. Moreover, repetition of the LZ transitions can induce quantum mechanical interference, which leads to Stueckelberg
or Ramsey-type oscillations.\cite{Stueckelberg,Ramsey} Since the theoretical scheme of observing LZ interference in qubits
was proposed by Shytov et al,\cite{Shytov} a series of beautiful experiments on LZ interference were implemented
in two level systems such as flux qubits\cite{Oliver, BernsPRL, Izmalkov2,Shevchenko3} and charge qubits,\cite{Sillanpaa} which provided an
alternative method to manipulate and characterize the qubit in the strongly driven regime.

A new regime of strong driving was reported in a recent work.\cite{Berns} Unlike previous experiments which employed a
two level system,\cite{Oliver,Sillanpaa,BernsPRL} the qubit in this experiment was driven through a manifold of several states
spanning a wide energy range. The
population of the qubit under large-amplitude fields exhibited a series of diamond-like interference patterns in the space
parameterized by flux detuning and microwave amplitude. The interference patterns, which displayed a multiscale character,
encoded the information of several energy levels of the system. In a recent work by Rudner et al., \cite{Rudner}
a skillful and concise method employing a
two-dimensional Fourier transform was used to study such a system. They transfer the measurement results of the energy domain
to the time domain, and realize a ``tomogram'' of the time evolution of the qubit phase, from which the decoherence
time of the qubit can be easily obtained.

In this article, rather than extract information by decomposing or translating the interference patterns, we start from the point
of reconstructing the interference patterns through analyzing the dynamics of the system, i.e., we map the time variable(including both
decoherence rate and field-driving rate)  onto the distribution of the qubit's population. Our model used to analyze multilevel
systems are based on the well developed theory of two-level systems.\cite{Shytov,BernsPRL,
Ashhab,Gefen,Shevchenko,Saitoo}
This article is organized as follows. In Sec.II the basic models were introduced firstly, from which we used rate equations to discuss
the dynamics of the two diamond-like interference patterns separately. The results agree with those of experiments\cite{Berns} very well. In Sec.III, we
discussed the effect of driving frequency and dephasing rate on the LZ interference. By tuning the driving frequency and  dephasing rate
one can expect different interference patterns. Some anomalous interference patterns, such as the moir$\acute{e}$-like pattern reported
in a recent experiment,\cite{Oliver} can be well explained using our model.

\begin{displaymath}
{ \textbf{II.\ MODEL AND METHOD }
}
\end{displaymath}

 We start from a driven two-level system subject to the effects of decoherence:\cite{BernsPRL,Ashhab}
 \begin{equation}
\hat{H}(t)=-\frac{\Delta}{2}\hat{\sigma}_{x}-\frac{h(t)}{2}\hat{\sigma}_{z},
\label{1}
\end{equation}
where $\Delta$ is the tunnel splitting; $\hat{\sigma}_{x}$ and $\hat{\sigma}_{z}$ are Pauli
matrices. $h(t)$ is the time dependent energy detuning from an avoided crossing:
 \begin{equation}
h(t)=\epsilon+Asin\omega t+\delta\epsilon_{noise}(t),
\label{2}
\end{equation}
where $\epsilon$ is the dc component of the energy detuning,  $A$ and $\omega$ are the amplitude
and frequency of the driving rf field respectively, $\delta\epsilon_{noise}(t)$ is the classical noise. As
discussed in ref [13], by using white noise model and perturbation theory, the
rate of LZ transitions between the states $\vert 0 \rangle$ and $\vert 1 \rangle$ can be obtained:
 \begin{equation}
W(\epsilon,A)=\frac{\Delta^{{2}}}{2}\sum_{n}\frac{\Gamma_{2}J^{2}_{n}(x)}{(\epsilon-n\omega)^{2}+\Gamma^{2}_{2}}
\label{3}
\end{equation}
where $\Gamma_{2}=1/T_{2}$ is the dephasing rate and  $J_{n}(x)$ are Bessel functions of the first
kind with the argument $x=A/\omega$. Eq.(\ref{3}) implies that the transition rate is proportional to
$\Delta^{2}$ which is decided by the energy structure of the system.

Extending Eq.(\ref{3}) to multilevel systems, the LZ transition rate between states $\vert i \rangle$ and
$\vert j \rangle$ can be written as:
 \begin{equation}
W_{ij}(\epsilon_{ij},A)=\frac{\Delta_{ij}^{{2}}}{2}\sum_{n}\frac{\Gamma_{2}J^{2}_{n}(x)}{(\epsilon_{ij}-n\omega)^{2}+\Gamma^{2}_{2}},
\label{4}
\end{equation}
where $ \Delta_{ij}$ is the avoided crossing between states $\vert i \rangle$ and $\vert j \rangle$, and
$\epsilon_{ij}$ is the dc energy detuning from the corresponding avoided crossing $\Delta_{ij}$.
Eq.({\ref{4}}) can be derived in the same way as in ref [13] considering only the direct coupling between states
$\vert i \rangle$ and $\vert j \rangle$. We emphasize that the situation will change for no decoherence, in which
coherence evolution among all coupled levels should be considered. While in our case, the coherence evolution is reduced to
rate equations as discussed below, which is appropriate in dealing with stationary population distribution.

\begin{figure}
\centering
\includegraphics[width=3.2375in]{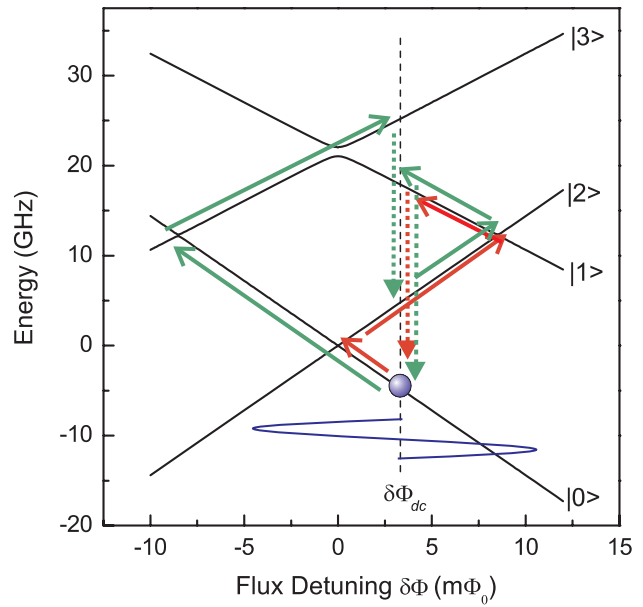}
\caption{ (color online) Energy-level diagram of a superconducting flux qubit illustrating the level-crossing positions for a
particular static flux detuning $\delta \Phi_{dc}$. Blue solid line represents the time-periodic detuning
$\Phi_{rf}sin\omega t$. The red path and the green path represent two different transition processes
leading to different interference patterns. The dotted lines represent intra-well relaxation processes.
}
\end{figure}
Hereafter, we focus on the multilevel superconducting flux qubit, a superconducting loop interrupted by three
Josephson junctions. Near flux bias $\Phi=0.5m\Phi_{0}$,  the system exhibits a double-well potential
parameterized by the flux detuning $\delta \Phi\equiv \Phi-0.5m\Phi_{0}$. Fig. 1 shows the lowest four levels
of the system as a function of flux detuning. Whenever the diabatic states $\vert i\rangle$ ($i$ =0, 1; right well with negative slope )
and $\vert j\rangle$ ($ j$ =2, 3; left well with positive slope) are degenerate, avoided crossing $\Delta_{ij}$ forms because
of the inter-well tunneling. If a microwave flux is applied (Fig. 1, blue sinusoid), the detuning flux of the system is
\begin {equation}
\delta \Phi(t)=\delta \Phi_{dc}+\delta \Phi_{ac}=\delta \Phi_{dc}+\Phi_{rf}sin\omega t, \label{5}
\end {equation}
with $\Phi_{rf}$, in the unit of flux, proportional to the microwave source voltage.
Then the time dependent energy detuning (see Eq.(\ref{2}))  of states $\vert i \rangle$ and $\vert j \rangle$
of the flux qubit can be expressed as:
\begin {equation}
h_{ij}(t)=(\vert m_{i}\vert+\vert m_{j}\vert)(\delta \Phi_{dc}+\Phi_{rf}sin\omega t)+\delta\epsilon_{noise}(t), \label{6}
\end {equation}
where $m_{i}=dE_{i}(\Phi)/d\Phi$ is the diabatic energy-level slope\cite{Berns} of state $\vert i\rangle$ in units of frequency per flux.
From Eq.(\ref{4}) and Eq.(\ref{6}), the LZ transition rate $W_{ij}$ is a function of $\delta \Phi_{dc}$ and $\Phi_{rf}$,
which can be easily controlled in experiments. To understand
simply, the tunnel splitting $\Delta_{ij}$ serves as a channel that connects different energy
states $\vert i \rangle$ and $\vert j \rangle$ when the systems are driven by large-amplitude fields.
Whether the channel is `open ' or `closed' is related with both the energy detuning $\epsilon_{ij}$ and the
amplitude of the driving field $A$.
As observed in experiments, the qubit population exhibited a series of `diamonds'
patterns\cite {Berns} in the space parameterized by $\Phi_{rf}$ and $\delta \Phi_{dc}$ . Here we mainly discuss the first two
diamonds produced by the transitions among the lowest four energy level states. The same methods can be used to
analyze other `diamonds'  resulting from transitions among higher energy levels. In order to have a clear physical picture of the interference
patterns, we analyze the first two diamonds in part A and part B separately, and make a more quantitative analysis in part C.

\begin{displaymath}
\textbf{A.\ THE FIRST DIAMOND }
\end{displaymath}

\begin{figure}
\centering
\includegraphics[width=3.4375in]{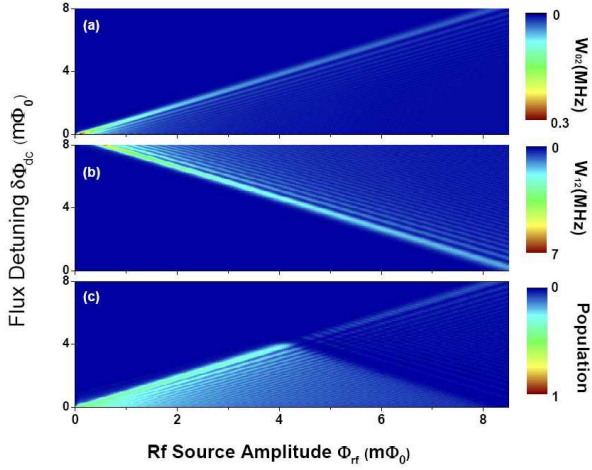}
\caption{(color online) (a) LZ transition rate $W_{02}$ versus flux detuning and driving amplitude. (b)LZ transition
rate $W_{12}$ versus flux detuning and driving amplitude. (c) Qubit population in the left well obtained from
Eq.(\ref{diamond1}). The left edge and the right edge mark the parameter values where the level crossings
$\Delta_{02}$ and $\Delta_{12}$ are first reached respectively.
The parameters we used are from experiments\cite{Berns} with driving frequency $\omega/2\pi=0.16$GHz,
$\Delta_{02}=0.013$GHz,$\Delta_{12}=0.09$GHz. The locations of the level crossings $\Delta_{02}$ and
$\Delta_{12}$ are 0m$\Phi_{0}$ and 8.4m$\Phi_{0}$ respectively. The diabatic energy-level slope $\vert
m_{0}\vert$($\vert m_{2}\vert$)=1.44GHz/m$\Phi_{0}$, and $\vert m_{1}\vert$($\vert m_{3}\vert$)
=1.09GHz/m$\Phi_{0}$. The dephasing rate we used in calculation $\Gamma_{2}/2\pi=0.05$GHz, and the
intra-well relaxation rate $\Gamma_{10}/2\pi=0.6$GHz.
}
\end{figure}
As shown in Figure 1, the red path represents the main transition process generating the first `diamond'. In order to describe the time evolution of the
population in the presence of driving fields, we employ a rate equation approach, in which the qubit level occupations
$p_{i}(i=0,1,2,3)$ obey:
\begin{align}
&\dot{p}_{0}=-p_{0}W_{02}+p_{1}\Gamma_{10}+ p_{2}(W_{02}+\Gamma_{20}),\nonumber\\
&\dot{p}_{1}=-p_{1}(W_{12}+\Gamma_{10})+p_{2}W_{12}, \nonumber\\
&p_{0}+p_{1}+p_{2}=1, \label{7}
\end{align}
where $W_{ij}$ is the LZ transition rate introduced in Eq.(\ref{4}), and $\Gamma_{ij}$ is the relaxation rate from
$\vert i \rangle$ to $\vert j \rangle$. It is noticed that some relaxation rates (e.g., $\Gamma_{01}$) that have little
effects on the results have been neglected for simplicity , and these rates would be considered in part C.
Moreover, we did not consider the population of $\vert 3\rangle$ for two reasons: (i) The driving-field amplitude is not large
enough to reach anti-crossing $\Delta_{03}$, resulting in no LZ transitions between $\vert 0 \rangle$ and $\vert 3 \rangle$.
(ii) In general, for the superconducting flux qubits, the intra-well relaxation rate $\Gamma_{10}$ is much larger than the LZ rate $W_{13}$. Therefore, the population of
$\vert 1\rangle$ mostly relaxed to $\vert 0\rangle$ rather than made a transition to $\vert 3 \rangle$. In part C these
approximations would be discussed in detail and we found they are valid for the reported experimental data.

 In the stationary case, $\dot{p}_{0}=\dot{p}_{1}=\dot{p}_{2}=0$. The qubit
population in the left well can be easily solved from Eq.(\ref{7}):
\begin {equation}
p_{2}=\frac{ W_{02}(W_{12}+\Gamma_{10}) }{ W_{12}(2W_{12}+W_{02}+\Gamma_{20})+\Gamma_{10}(2W_{02}
+W_{12}+\Gamma_{20}) }. \label{diamond1}
\end {equation}
Fig.2(c) is the contour plot according to Eq.(\ref{4}) and Eq.(\ref{diamond1}) with the parameters from experiments\cite{Berns}.
The main features of the theoretical results agree with the experimental data very well. Since the pattern is symmetric about 0 detuning,
we only plot half of the pattern. Qualitatively, using Eq.(\ref{diamond1}), we split the space into three regimes:

$p_{2}=\left\{ \begin{array}{ll} 0 & \textrm{$W_{02}$ is off}\\
\frac{W_{02}}{2W_{02}+\Gamma_{20}} & \textrm{$W_{02}$ is on;$W_{12}$ is off}\\
\frac{W_{02}}{W_{12}} \to 0 & \textrm{$W_{02}$ is on; $W_{12}$ is on}	.
\end{array} \label{three regime} \right .$\\

When the amplitude of rf field is small, $W_{02}$ is off.
The channel that connects states $\vert 0\rangle$ and $\vert 2\rangle$ is cut off, resulting in no population transfer to the
left well. The minimum amplitude required to switch $W_{02}$ on is proportional to the detuning. Therefore, we observed an edge with
positive slope at left. With the amplitude increased, we may enter a regime where $W_{02}$ is on and $W_{12}$ is off. The qubit
serves as a two level system with the channel to higher energy levels cut off. Since the minimum amplitude to switch on the $W_{12}$
linearly decreases with the detuning, we saw a shadow with negative slope on the right side.  There is another regime where both $W_{02}$
and $W_{12}$ are on. Because $W_{12}$ is much larger than $W_{02}$, the population
of $\vert1\rangle$ is more likely to be excited to $\vert 2\rangle$, and then relaxes back to $\vert 0\rangle$, thus suppressing
the net population transfer to the left well (see the red path in Fig.1). In addition, the population on the peaks of the interference
patterns should be less than 0.5 because of the spontaneous relaxation to the ground state. This was confirmed in the experiments.

\begin{figure}
\centering
\includegraphics[width=3.4375in]{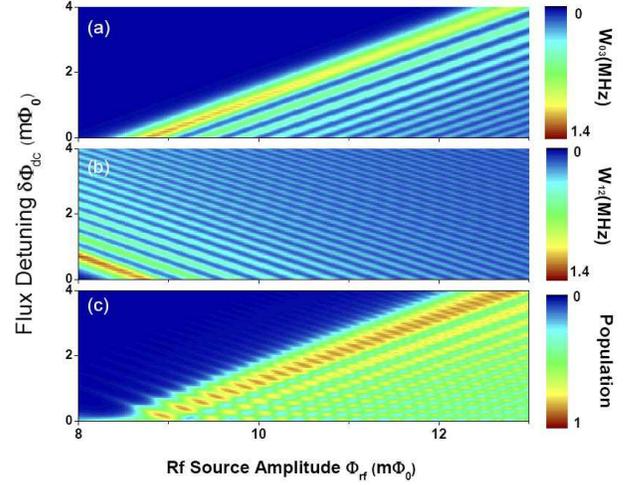}
\caption{(color online) (a) LZ transition rate $W_{03}$ versus flux detuning and driving amplitude. (b)LZ transition rate $W_{12}$
versus flux detuning and driving amplitude. (c) Qubit population in the left well obtained from Eq.(\ref{11}).
Features of population inversion and checkerboard pattern are notable. The parameters we used are the same
with those in Fig.2. }
\end{figure}
\begin{displaymath}
\textbf{B.\ THE SECOND DIAMOND }
\end{displaymath}
The `second-diamond' interference patterns correspond to the transition processes shown in Fig.1(green path).
Using a similar method with that of part A and neglecting weak transitions, we can write the rate equation for the
population evolution as:
\begin{align}
&\dot{p}_{0}=-p_{0}W_{03}+p_{1}\Gamma_{10}+p_{2}\Gamma_{20}+p_{3}W_{03},\nonumber\\
&\dot{p}_{1}=-p_{1}(W_{12}+\Gamma_{10})+p_{2}W_{12}, \nonumber\\
&\dot{p}_{2}=p_{1}W_{12}-p_{2}(W_{12}+\Gamma_{20})+p_{3}\Gamma_{32},\nonumber\\
&p_{0}+p_{1}+p_{2}+p_{3}=1, \label{10}
\end{align}
where $W_{ij}$ and $\Gamma_{ij}$ have the same definitions as those in part A. In the stationary case
$\dot{p}_{0}=\dot{p}_{1}=\dot{p}_{2}=\dot{p}_{0}=0$, Eq.(\ref{10}) can be analytically solved.
Considering $\Gamma_{10}$, $\Gamma_{32}\gg W_{12}$,$W_{03}$, which is the case
in experiments, the population in the left well can be obtained in a very simple form:
\begin{equation}
p_{L}=p_{2}+p_{3}\backsimeq\frac{W_{03}}{W_{03}+W_{12}+\Gamma_{20}}. \label{11}
\end{equation}
Fig.3(c) is the contour plot using Eq.(\ref{11}) with experimental parameters. The results, which show
notable features of both checkerboard patterns and population inversion, agree with those of the experiments very well.
The population inversion is a straightforward result of Eq. (10). When $W_{03}$ is on and $W_{12}$ is off, since $W_{03}>\Gamma_{20}$,
the system forms a $\lambda$-type three-level structure. $p_{L}$ is larger than 0.5 and even close to 1. Actually, one can use this property
to generate controllable population inversion to produce lasers. However, once
$W_{12}$ is on, the population of $\vert 2\rangle$ is pumped to $\vert 1\rangle$ and relaxes to $\vert 0\rangle$
rapidly. Because $W_{12}>W_{03}$, $p_{L}<0.5$, thus destroying the population inversion(see the green path in Fig.1).

\begin{figure}
\centering
\includegraphics[width=3.4375in]{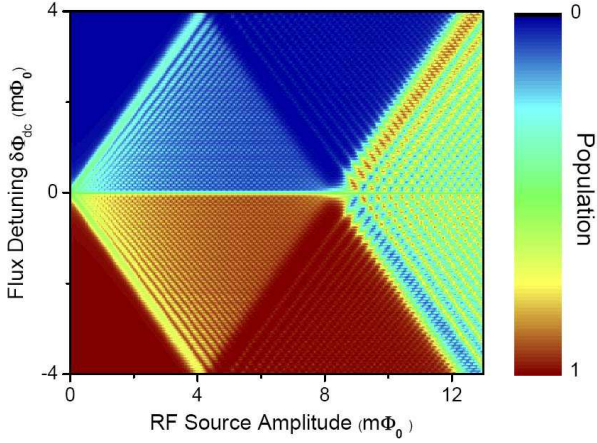}
\caption{(color online) Calculated qubit population versus flux detuning and driving amplitude. The first diamond
corresponds to the red path and the second diamond corresponds to the green path in Fig.1. The
parameters we used in calculation are: the inter-well relaxation rate $\Gamma_{20}/2\pi=0.05$MHz,
the avoided crossing $\Delta _{13}$=0.5GHz, and the temperature $T$ used in $\Gamma_{02}=
\Gamma_{20}$ exp$(-E_{02}/k_{b}T)$ is 20mK. Other parameters are the same with those in Fig.2.  }
\end{figure}

\begin{figure}
\centering
\includegraphics[width=3.4375in]{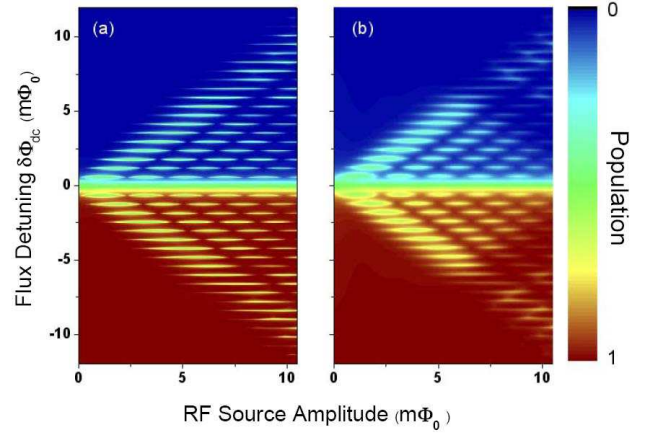}
\caption{(color online) (a) Qubit population obtained from
Eq.(\ref{12}) with driving frequency $\omega/2\pi=1.2$GHz. Moir$\acute{e}$-like patterns reveal
for resonances $n>12$. The parameters we used are from experiments\cite{Oliver,Yu}, with
$\Delta_{02}=0.004$GHz and $\Delta_{12}$ is fitted to be 0.09GHz. The locations of the level
crossings $\Delta_{02}$ and $\Delta_{12}$ are 0m$\Phi_{0}$ and 13.1m$\Phi_{0}$ respectively.
The diabatic energy-level slope $\vert m_{0}\vert$($\vert m_{2}\vert$)=1.01GHz/m$\Phi_{0}$,
and $\vert m_{1}\vert$($\vert m_{3}\vert$)=0.91GHz/m$\Phi_{0}$. The dephasing rate we used in
calculation $\Gamma_{2}/2\pi=0.05$GHz, the inter-well relaxation rate $\Gamma_{20}/2\pi=0.05$
MHz, and the intra-well relaxation rate $\Gamma_{10}/2\pi=0.6$GHz. (b) The dephasing rate is
increased to $\Gamma_{2}/2\pi=0.2$GHz, leading to a diamond-like interference pattern, in which the
area of missing parts increases.}
\end{figure}

\begin{displaymath}
 \textbf{C.\ COMBINE THE TWO DIAMONDS }
\end{displaymath}
Having addressed the two diamonds separately, now we can make a more integral and quantitative
study of the interference patterns. Here the driving-field amplitude is increased from $0$ to a large value which has
reached $\Delta_{03}$ and $\Delta_{12}$ crossings. In this situation, all the transition processes in Fig.1 should be
considered, and the rate equation can be written as:
\begin{align}
&\dot{p}_{0}=-p_{0}(W_{02}+W_{03}+\Gamma_{02})+p_{1}\Gamma_{10}+p_{2}(W_{20}+\Gamma_{20})\nonumber\\
&\phantom{\dot{p}_{0}=}+p_{3}W_{03},\nonumber\\
&\dot{p}_{1}=-p_{1}(W_{12}+W_{13}+\Gamma_{10})+p_{2}W_{12}+p_{3}W_{13}, \nonumber\\
&\dot{p}_{2}=p_{0}(W_{02}+\Gamma_{02})+p_{1}W_{12}-p_{2}(W_{02}+W_{12}+\Gamma_{20})\nonumber\\
&\phantom{\dot{p}_{2}=}+p_{3}\Gamma_{32},\nonumber\\
&p_{0}+p_{1}+p_{2}+p_{3}=1, \label{12}
\end{align}

\begin{figure*}
\centering
\includegraphics[width=6.4375in]{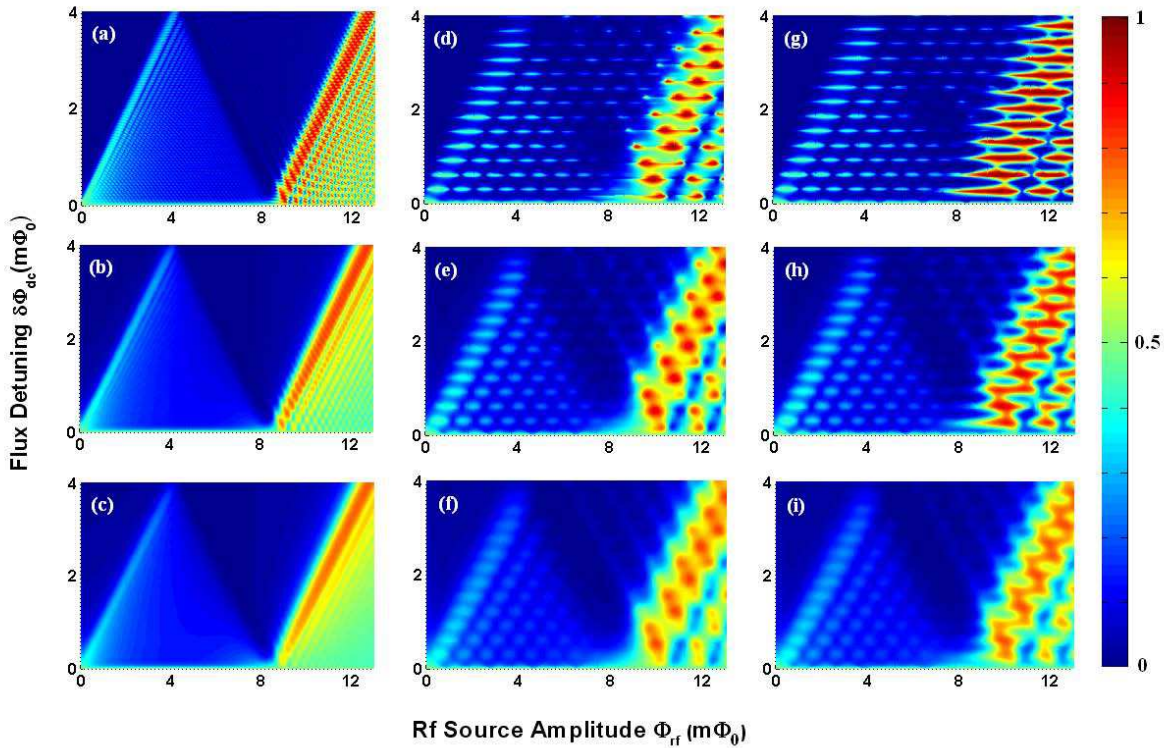}
\caption{(color online) Qubit population obtained from
Eq.(\ref{12}) with driving frequency $\omega/2\pi=0.16$GHz, 0.879GHz and 0.8886GHz from left to
right, and dephasing rate $\Gamma_{2}/2\pi$=0.05GHz, 0.2GHz and 0.4GHz from top to bottom,
respectively. The parameters of the superconducting flux qubit are the same with those in Fig.2. }
\end{figure*}

which can also be solved analytically. Since the analytical solution is too complex to extract a physical picture,
we did not write it out explicitly here. Nevertheless, we solved Eq. (\ref{12}) numerically and obtained the
qubit population as functions of the amplitude and flux detuning as shown in Fig.4.
The agreement between our results and those of the experiments is remarkable. The right edges of the second diamond are not shown in Fig.4 because it requires
the presence of higher excited states, which could be considered easily with the method we
discussed here.

In addition, there is no explicit signature of coherent traversal through avoided
crossing $\Delta_{13}$ observed in experiments\cite{Berns}. In order to clarify this point,
we changed the value of $\Delta_{13}$ from 0 to 0.8GHz. The
interference patterns obtained from Eq.(11) exhibit almost no change, indicating that the LZ transitions at
$\Delta_{13}$ has negligible contribution to the whole process.
Moreover, we have compared the results of part C to those of part A and B
respectively. They accorded very well, indicating that our approximation in part A and B is appropriate.

\begin{center}
\textbf{III.\ EFFECT OF DRIVING-FREQUENCY AND DEPHASING RATE}
\end{center}

As discussed in ref [12], subsequent LZ-tunneling events can interfere only when phase coherence is preserved, requiring that
the time interval of subsequent LZ tunneling events $\tau_{p}$ must be less than decoherence
time $\tau_{coh}$ of the system ($\tau_{p} < \tau_{coh}$ ). Since $\tau_{p} \varpropto \omega^{-1}$, if the
qubit's decoherence time $\tau_{coh}$ is short due to strong coupling to environment, a large driving frequency is required
to observe LZ interference. Therefore, it is of great meaning to discuss the behavior of the qubit
driven by high-frequency fields. We emphasize that the high frequency we discussed here is still in the regime dominated by LZ
transitions, which means that the analysis in Sec.II is still valid.

Another advantage of using high frequency is that one can resolve more features in the interference patterns.
At low frequency, the microwave can match the energy level spacing approximately at all flux detuning, corresponding to the almost continuous
band of $W_{ij}$ in Fig.2 and Fig.3.
Actually, these continuous bands are composed of resonant peaks (with single and multiple photon effect) that are overlapped with each other.
The distance between these peaks is the driving frequency.
At high frequency, the distance of these overlapped resonances increased and the continuous band in Fig 2. and Fig. 3 become
a series of discrete peaks.
This was demonstrated by recent experiments. By using higher driving frequency, clear LZ interference pattern is observed
and the results show some new characters(e.g., partly missing
fringes) that are different from the `diamond' interference patterns\cite{Berns}.
We found that $\Gamma_{2}\backsimeq \omega/2\pi $ is the critical situation that the interference patterns transform between moir¡äe-like
and diamond-like types.
To explain the moir$\acute{e}$-like pattern\cite{Jain}
observed for resonances $n >12$, where $n$ represents the $n$-photon resonances, higher excited
states need to be considered. By inserting the experimental parameters\cite{Yu, Oliver} into Eq.(11), the qubit
population was obtained (see Fig.5(a)). The calculated results, with some fringes partly missing and some fringes intact,
agree with the experiments very well.\cite{Oliver}
These features can be easily understood using Eq. (9). The missing part of the fringes at $n>12$ corresponds where both $W_{02}$ and $W_{12}$ are on.
The intact fringes (e.g., n=14) correspond where $W_{02}$ is on while $W_{12}$ is off.

For higher dephasing rate, the width of the resonance peaks for all $W_{ij}$ will increase. As a result, the area of the region
where both $W_{02}$ and $W_{12}$ are on also expands, resulting in an increasing area of missing fringes(see Fig. 5(b)).
Similar to that of the part A, the missing fringes form a shadow with negative slope.

We investigated the effect of driving frequency and dephasing rate on the LZ interference. Shown in fig.6 are some examples.
With dephasing rate increasing, the individual resonances are no longer distinguishable and merged into a continuous band (Fig.6 (d-f)).
However, by increasing the driving frequency to $\omega/2\pi > \Gamma_{2}$, we rebuild the interference again (Fig.6(d,g)),
as discussed in ref[13]. It is interesting that there are some new features emerging during these processes.

As we discussed in Sec. II, whenever $W_{03}$ and $W_{12}$ are on simultaneously, the population will be pumped back and a missing part
will be generated on the fringes (Fig. 3). For small dephasing rates, the resonant peaks are very sharp.
The resonant peaks are difficult to hit the spots where both $W_{03}$ and $W_{12}$ are on. We thereby saw high quality moire-like
interference patterns.  However, for large dephasing rates, the resonant peaks are very wide along the axis of flux detuning.
At many places $W_{03}$ and $W_{12}$ manipulate the resonant peaks simultaneously and we observed complicate patterns that are mainly
diamond-like. Similar anomalous pattern was reported in recent experiments\cite{Izmalkov2}.

\begin{center}
\textbf{IV.\ CONCLUSION}
\end{center}

We have analyzed the interference patterns observed in recent experiments
which demonstrated an innovative approach to make spectroscopic measurement of a quantum system.
The analytical results obtained from our model are agreed with the experiments very well, and the method can be extended to
investigate the response of higher energy levels under a large-amplitude driving field. Moreover, we investigated
the effect of driving frequency and dephasing rate on the LZ interference.  The fast dephasing will destroy the interference.
However, it is possible to overcome the dephasing by increasing rf frequency.
We have not only explained the moir$\acute{e}$-like pattern observed in a recent experiment, but also discussed some interesting anomalous
patterns that could be expected in experiments. Our results can be used to understand the LZ interference in multilevel
system under different driving frequencies and decoherence rates. The model and method used in this article can be extended
to other systems with multiple energy levels structure.

In addition, we noticed a recent work\cite{Lupascu} in which microscopic two-level systems (TLS) were observed in flux qubits by using one- and
two-photon spectroscopy method. We proposed that it is very convenient to detect such TLSs with the amplitude spectroscopy method demonstrated
by Berns et al,\cite{Berns} and the analytical methods in this article will be still valid in such case.\\

\begin{center}
\textbf{V.\ ACKNOWLEDGEMENT}
\end{center}

We thank M. S. Rudner and W. D. Oliver for useful discussions.
This work was partially supported by NSFC (10674062, 10725415), the
State Key Program for Basic Research of China (2006CB921801), and the
Doctoral Funds of Ministry of Education of the People's Republic of
China (20060284022).\\

\end{document}